\documentclass[a4paper,11pt]{article}
\usepackage{pos}
\usepackage{float}

\title{Next-day observation processing for the LST-1 and MAGIC}
\ShortTitle{Next-day, LST-1 and MAGIC}

\author*[a]{A. Dinesh}
\author[a]{M. Láinez}
\author[a]{R.A. Cerviño}
\author[b]{D. Morcuende}
\author[c]{A. Moralejo}
\author[a] {J.L. Contreras}
\author[d]{A. Baquero}
\author[e]{J. Lozano}
\onbehalf{for the MAGIC Collaboration}

\affiliation[a]{IPARCOS-UCM, Instituto de Física de Partículas y del Cosmos, and EMFTEL department, Universidad Complutense de Madrid,\\
  Plaza de Ciencias 1, Madrid, Spain}

\affiliation[b]{CTAO Project Office, Deutsches Elektronen-Synchrotron (DESY),\\
D-15738 Zeuthen, Germany }

\affiliation[c]{Institut de Fisica d’Altes Energies (IFAE), The Barcelona Institute of Science and Technology,\\
Campus UAB, 08193 Bellaterra (Barcelona), Spain}

\affiliation[d]{Faculty of Science and Technology, Universidad del Azuay,\\
Cuenca, Ecuador}

\affiliation[e]{Departamento de Physics and Mathematics, University of Alcalá UAH,\\
Pza. San Diego, 28801, Alcalá de Henares, Madrid, Spain}

\emailAdd{adinesh@ucm.es}
\emailAdd{malainez@ucm.es}

\onbehalf{on behalf of the CTAO-LST project and the MAGIC collaboration}

\abstract{
The MAGIC and LST-1 telescopes, located at the Roque de los Muchachos Observatory on La Palma, operate dedicated On-Site Analysis (OSA) pipelines that provide rapid, automated processing of observational data. These systems produce high-level data products just a few hours after observations are completed, enabling quick-look analyses, next-day data quality assessments, and rapid-response science such as flare detection and Target of Opportunity follow-ups. OSA pipelines have been in continuous operation since 2012 for MAGIC and since 2021 for LST-1, automatically processing nightly data using the standard analysis chain. The experience gained from both systems provides essential lessons for the development of Cherenkov Telescope Array Observatory’s (CTAO’s) on-site analysis, demonstrating the practical and scientific benefits of fast data processing in Cherenkov telescopes.
}

\ConferenceLogo{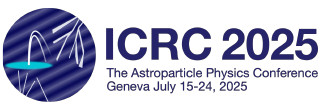}

\FullConference{39th International Cosmic Ray Conference (ICRC2025)\\
 15–24 July 2025\\
Geneva, Switzerland\\}

\begin{document}
\maketitle

\section{Introduction}
\begin{figure}[htbp]
    \centering
    \includegraphics[width=0.45\textwidth]{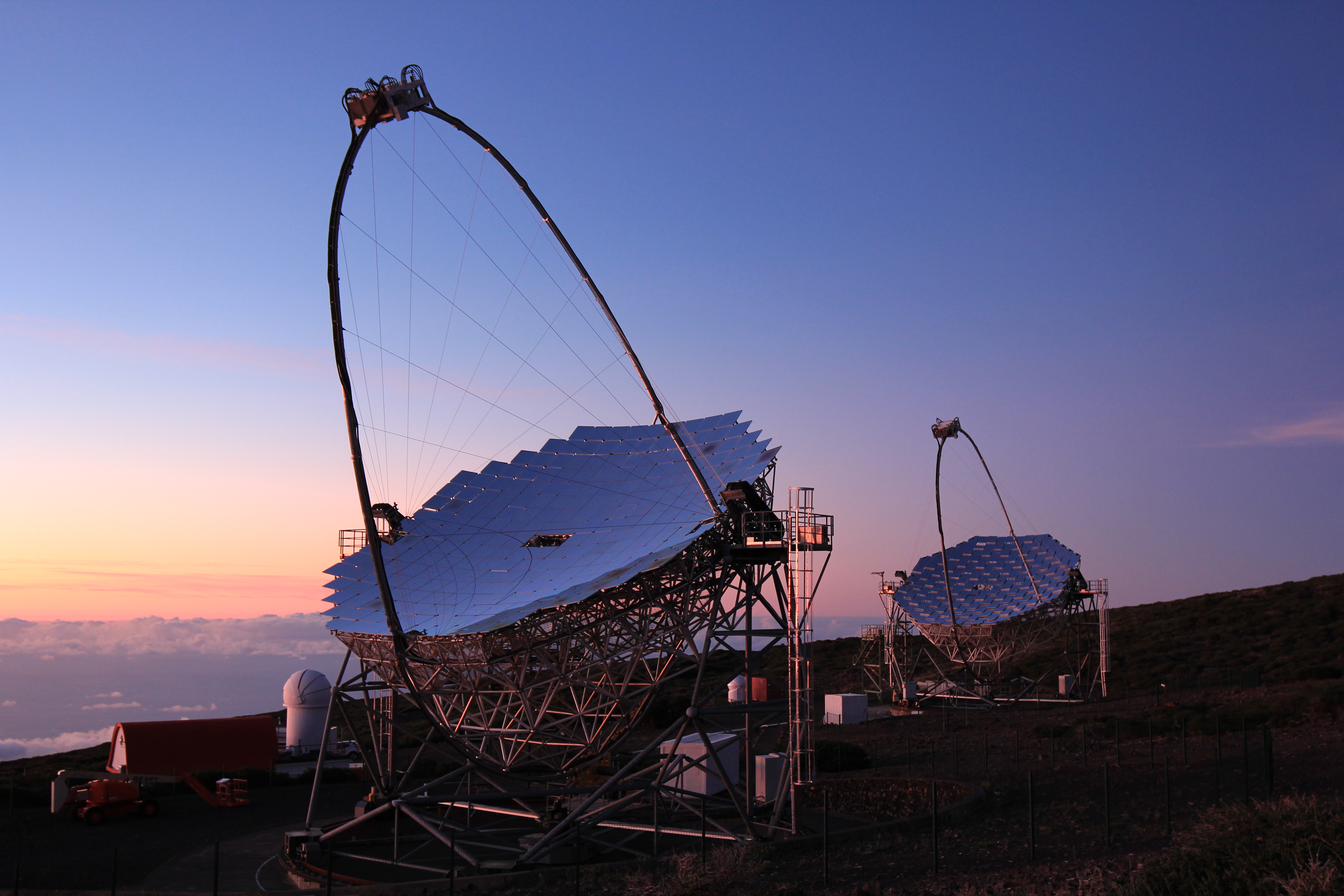}
    \hfill
    \includegraphics[width=0.45\textwidth]{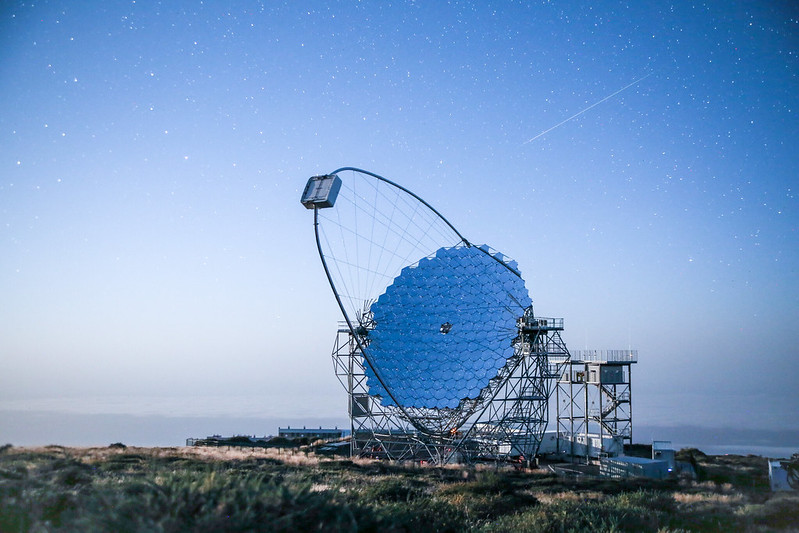}
    \caption{The MAGIC stereo system (left) \cite{Photo-Ceribella} and the LST-1 telescope (right) \cite{Photo-Inada}, both located at the Observatorio del Roque de los Muchachos in La Palma. These imaging atmospheric Cherenkov telescopes are part of the current and next-generation facilities for very-high-energy gamma-ray astronomy.}

    \label{fig1}
\end{figure}

The first CTAO Large-Sized (LST-1) and the MAGIC telescopes are two Imaging Atmospheric Cherenkov Telescope (IACT) systems located at the Roque de los Muchachos Observatory on La Palma, Canary Islands, Spain. Figure \ref{fig1} shows both instruments. Rapid availability of reliable analysis products to both collaborations is crucial for science alerts, scheduling, and data quality assurance. Additionally, the large size of the raw data makes its transmission to off-site data centres within a short time window challenging.
On-Site Analysis (OSA) pipelines have been operational for MAGIC since 2012 and for the LST-1 since 2021. Both pipelines automatically analyse data using the standard analysis chain. These data serve as input for the daily checks performed on the data and the rapid analysis activities. In the case of MAGIC, an automatic analysis chain also derives the detection significance, spectra, light curves and sky maps of the observations. These last steps of the analysis are also currently being included into the LST-1 pipeline.

\section{Context}

MAGIC is a two-telescope stereo system. With a trigger rate of $\sim 300$ evts/s, the telescopes record up to 2 Terabytes (TB) of raw data per day \cite{MAGIC-perf}. The OSA pipeline processes data up to high-level products, such as sky maps and energy spectra. It operates on a dedicated system comprising approximately 50 cores and 40 TB of disk storage. 

The LST-1 is a single telescope. With a trigger rate of $\sim 7-8$ kevts/s, it produces up to 30 TB of raw data per day \cite{LST1-perf}. Its OSA pipeline processes raw to reconstructed events (DL2, see Section \ref{sec:structure}), utilising an on-site data centre of 1,800 cores and more than 5 Petabytes (PB) of disk storage.

Both observatories rely on fast real-time analyses, which provide immediate estimations during the data taking at the expense of reduced sensitivity. They differ from the next-day analysis described here, which utilises the complete chain and provides high-quality data. Both approaches complement each other. Results from the next-day analysis are usually available before 14:00 local observatory time.

\section{Structure of the systems}
\label{sec:structure}
\begin{figure}[htbp]
    \centering
    \includegraphics[width=0.8\textwidth]{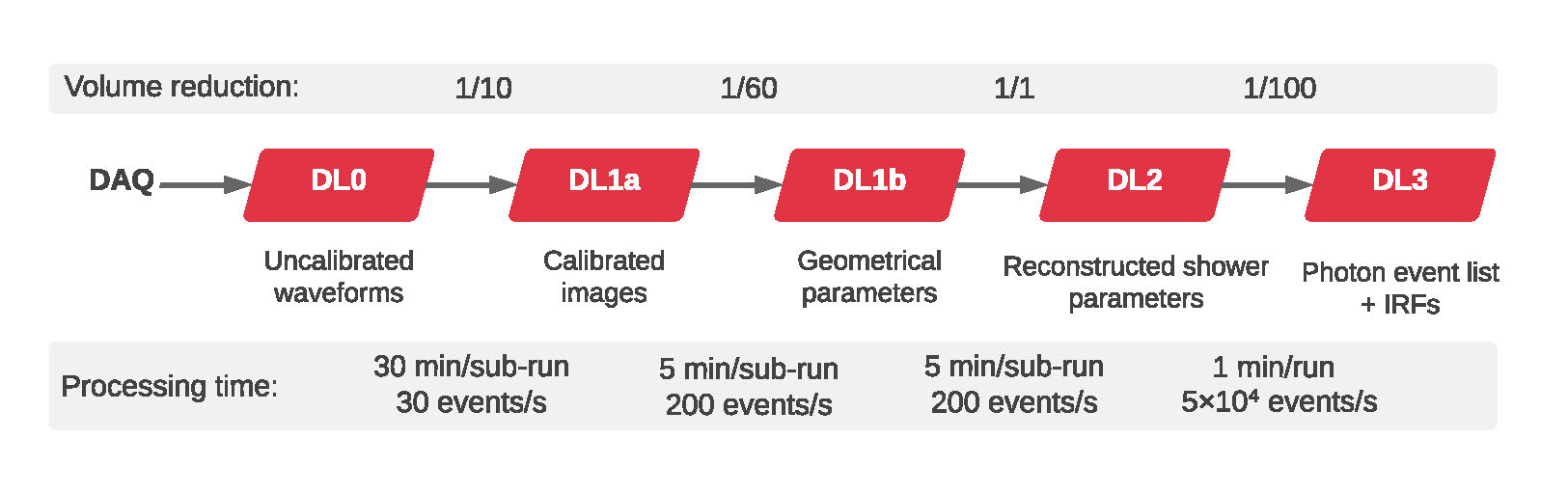}
    \caption{A simplified view of the LST-1 data flow, data size and processing time for each analysis step, it is very similar to the MAGIC one. Taken from reference \cite{Tesis-DMorcuende}.}
    \label{fig2}
\end{figure}

The data analysis of both observatories is similar. The telescopes are equipped with fast photomultiplier (PMTs)  cameras with around 1000 (MAGIC) or 2000 (LST-1) pixels. When the Cherenkov light from an Extensive Air Shower, or a calibration pulse, arrives at the camera, the PMTs produce fast electronic pulses. After the trigger occurs, the pulse is digitised in a readout window, with 80 (MAGIC) or 40 (LST-1) samples per pixel. This data is recorded together with associated metadata and constitutes the lowest level data: Data Level 0 (DL0). After performing calibration, the reconstructed charge and arrival time per pixel are obtained (DL1). Images are then reconstructed, and the physical parameters of the original particle are estimated (DL2). Photon candidates are then selected in event lists, and the Instrument Response Functions are produced (DL3). This is the data that will be delivered to users in the future and serves as the basis for higher-level products. These steps are represented in Figure \ref{fig2}.

The OSA system in the LST-1 is an evolution of the MAGIC OSA one; they are based on the same scheme, but the code has been almost completely rewritten. 
MAGIC OSA's pipeline is based on the Mars package \cite{Mars} while the LST-1 OSA pipeline uses the lstchain software \cite{lstchain}. Several individuals have contributed to the work over the years (e.g., \cite{Tesis-DMorcuende}, \cite{Tesis-ABaquero}, \cite{Tesis-MLainez}). Both systems are primarily written in Python. Crontab jobs launch the main scripts, and the existence of key files controls their flow. In the pre-processing phase, information on observations is obtained from the DAQ databases and the slow control files, which contain the pointing information from the telescope. To initiate the analysis, a summary of the daily observations is compiled to guide the process, and the analysis jobs are subsequently launched based on it. Dependencies between different steps of the processing are taken into account when the jobs are prepared and during their execution.

During the first steps of the processing phase, which is the most time-consuming, the analysis is parallelised to the lowest possible level. Data is split into sets of files, equivalent to approximately 10 seconds of observation for LST-1 or 20 minutes for MAGIC. Individual jobs are launched for these units using a batch system (Torque for MAGIC and SLURM for LST-1). In the LST-1, several hundred cores are used in parallel, while in MAGIC the number of concurrent jobs is of the order of 10-20, albeit longer ones. During the processing, OSA interacts periodically with the system to check which jobs are finished and tag those with problems. The following steps are also run on the batch system but with a lower level of parallelisation.

LST-1 OSA integrates the production of data check summaries for the day and their transfer to websites where experts can later review them. In the case of MAGIC, the data check is performed by an independent system. In both cases, after the processing is finished, different processes merge the files, move them to their final locations and clean up the used directories.

After the processing is finished, a different system, independent of OSA, handles the transfer of data to the off-site data centres, where it becomes accessible to users. There, it is also available for rapid dedicated analysis, such as those conducted by flare advocates or triggered by "Targets of Opportunity".

\begin{figure}[htbp]
    \centering
    \includegraphics[trim=0 5 0 0, clip,width=0.8\textwidth]{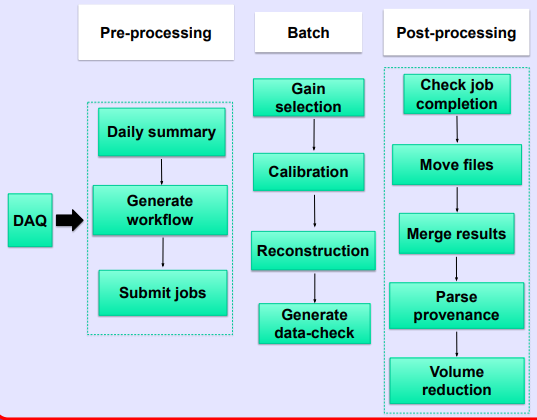}
    \caption{A simplified view of the LST-1 data flow, very similar to the MAGIC one.}
    \label{fig3}
\end{figure}

A simplified scheme of the LST-1 OSA flow is represented in Figure \ref{fig3}.

\begin{figure}[htbp]
    \centering
    \includegraphics[width=0.8\textwidth]{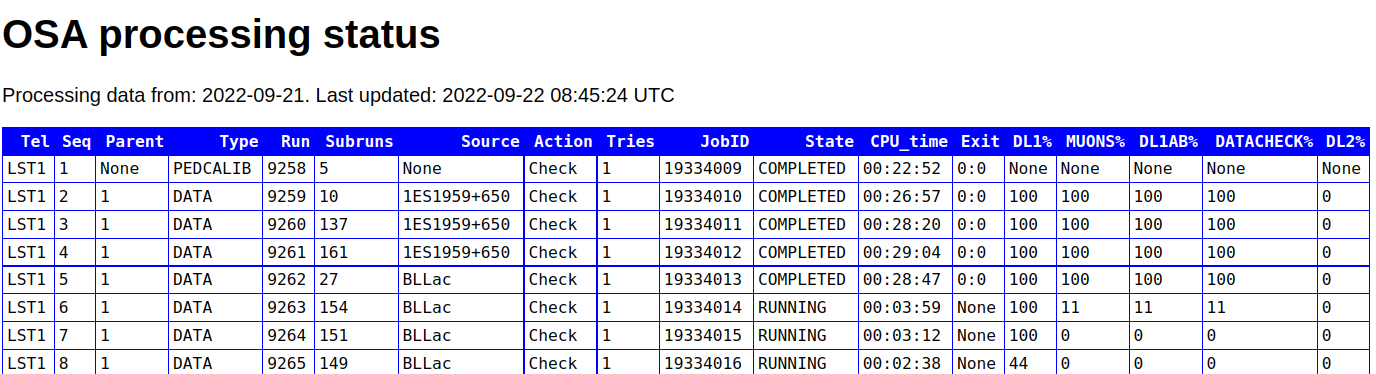}
    \caption{Screenshot of one of the web pages that shows the LST-1 OSA data analysis status daily. The page is refreshed every ten minutes. Taken from reference \cite{Tesis-DMorcuende}}
    \label{fig4}
\end{figure}

OSA operates automatically, but several tools allow for the supervision of its functioning. Web pages, such as the one shown in Figure \ref{fig4}, display the status of the daily analysis. They are refreshed every few minutes while the analysis is running. Through them, the OSA teams and the whole collaboration can easily check the progress of the data reduction. Additionally, in the event of non-recoverable errors, the scripts notify the OSA teams via email. 

To avoid both deadlocks and infinite loops, OSA incorporates an error-check system. When a process fails, it is relaunched a certain number of times, as its failure can be due to temporary conditions. After a few trials, the process is stopped and declared vetoed. Also, a limited amount of error recovery has been implemented to solve the most common issues. This information, including the number of trials for each job and any vetoing that may have been applied, is reflected in the above-described web pages. The OSA error recovery procedures are currently only available in MAGIC, but they are planned to be implemented for the LST-1 as well.

\begin{figure}[H]
    \centering
    \includegraphics[trim=6 0 0 0, clip,width=0.45\textwidth]{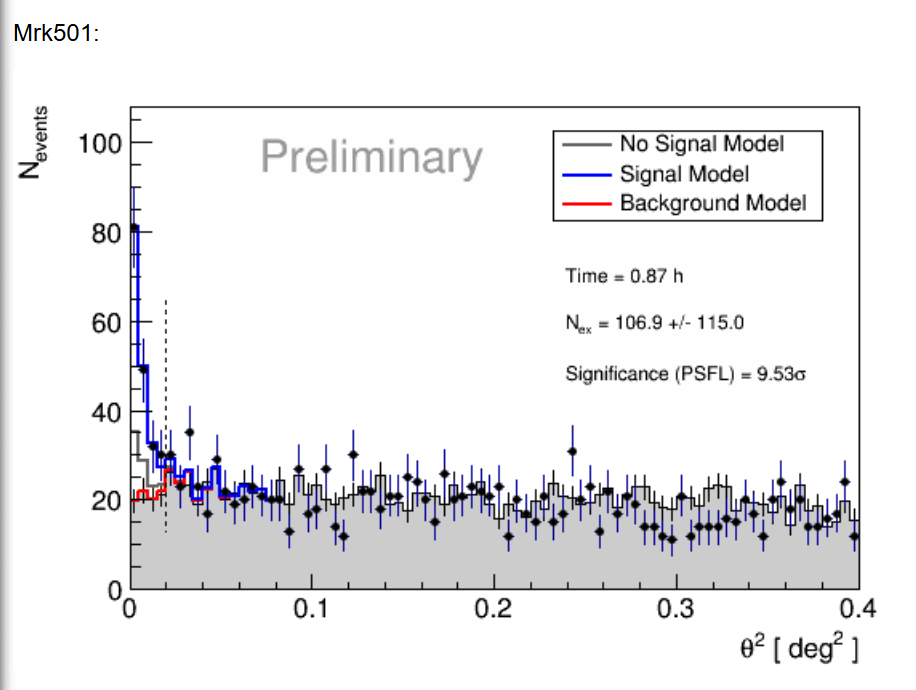}
    \hfill
    \includegraphics[trim=7 0 0 0, clip,width=0.45\textwidth]{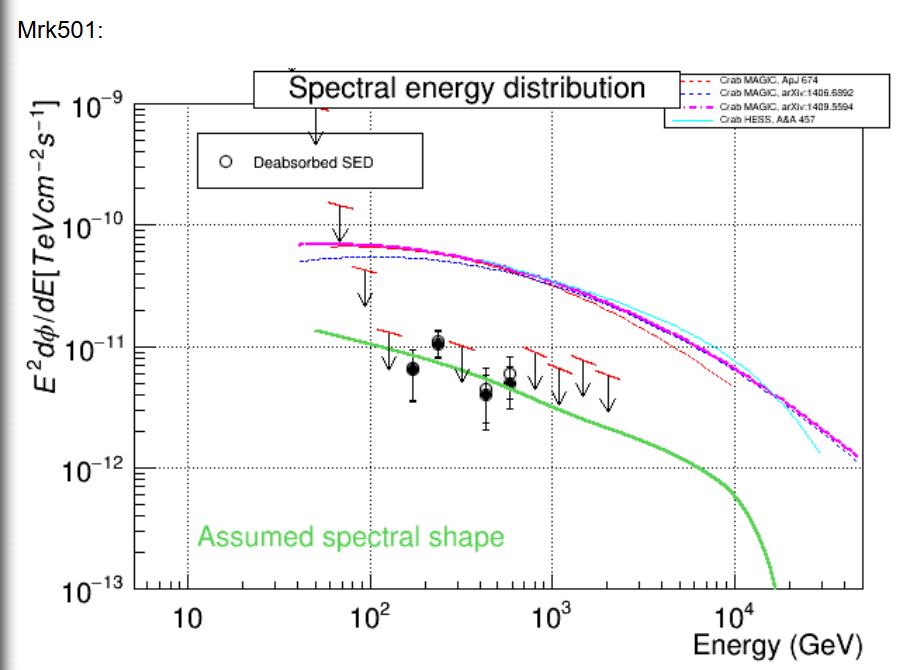}
    \caption{Examples of high-level plots produced by the MAGIC OSA pipeline from a recent observation of Mrk 501. \textbf{Left:} Classical $\theta^2$ plot showing the significance of the source detection. \textbf{Right:} Spectral energy distribution (SED) corrected for extragalactic background light (EBL) absorption. These plots are generated automatically every day and are publicly available online for the collaboration members.}
    \label{fig5}
\end{figure}

As previously commented, in the case of MAGIC, OSA processing extends beyond that of the LST-1, producing not only lists of photon candidates but also higher-level products, such as significance plots, sky maps, and spectra. Figure \ref{fig5} shows a so-called $\theta^2$ plot for a recent observation and a tentative SED (Spectral Energy Distribution). The first represents the histogram of events as a function of distance to the source, for both signal and background regions. Using these distributions, along with an estimate of the observatory's point spread function (PSF), the significance of a possible detection in an area close to the source position is estimated. The second plot contains a fast estimation of the source SED based on approximated instrument response functions.

\section{Lessons learned}
From the development and operation of the OSA systems over several years, we can extract several lessons that can be applied to similar observatories, in particular, the future CTAO ones.

As a first lesson, we are convinced that although disposing of a complete next-day analysis is a challenge, it is also a necessity for a Very-High-Energy (VHE) observatory.  Firstly, it provides fast access to processed data, enabling a quick response to transient events. Additionally, it detects hardware and software problems that inevitably arise during telescope operation and facilitates the implementation of solutions within the analysis chain.

In the second place, it is clear that the next-day pipelines can not be isolated from the rest of the processing but should be tightly coordinated with the observatory databases and simulation tools.

As the third one, we note that although all tasks should be automatic, means for easy human intervention are always needed, and a comprehensive scheme for error detection and recovery should be established from the initial design.

These are only the main general points; several more detailed ones are detailed in references \cite{Tesis-DMorcuende}, \cite{Tesis-ABaquero} and \cite{Tesis-MLainez}. 

\section{Conclusions}
On-Site Analyses have been working continuously for 13 years in MAGIC and 5 years in the LST-1 in the challenging environment of a mountain observatory.

Simple in design and effective, they have contributed to the success of both collaborations,  providing fast, high-quality data, data check measures and, for MAGIC, quick basic physics results.
 
They can serve as a prototype, test bed and check for the future CTAO onsite pipelines.

\vspace{0.2cm}
\textbf{Full Author List: CTAO-LST Project}

\tiny{\noindent
K.~Abe$^{1}$,
S.~Abe$^{2}$,
A.~Abhishek$^{3}$,
F.~Acero$^{4,5}$,
A.~Aguasca-Cabot$^{6}$,
I.~Agudo$^{7}$,
C.~Alispach$^{8}$,
D.~Ambrosino$^{9}$,
F.~Ambrosino$^{10}$,
L.~A.~Antonelli$^{10}$
% ...
The CTAO-LST Project
}\\

\tiny{\noindent$^{1}${Department of Physics, Tokai University, 4-1-1, Kita-Kaname, Hiratsuka, Kanagawa 259-1292, Japan}.
$^{2}${Institute for Cosmic Ray Research, University of Tokyo, 5-1-5, Kashiwa-no-ha, Kashiwa, Chiba 277-8582, Japan}.
$^{3}${INFN and Università degli Studi di Siena, Dipartimento di Scienze Fisiche, della Terra e dell'Ambiente (DSFTA), Sezione di Fisica, Via Roma 56, 53100 Siena, Italy}.
$^{4}${Université Paris-Saclay, Université Paris Cité, CEA, CNRS, AIM, F-91191 Gif-sur-Yvette Cedex, France}.
$^{5}${FSLAC IRL 2009, CNRS/IAC, La Laguna, Tenerife, Spain}.
$^{6}${Departament de Física Quàntica i Astrofísica, Institut de Ciències del Cosmos, Universitat de Barcelona, IEEC-UB, Martí i Franquès, 1, 08028, Barcelona, Spain}.
$^{7}${Instituto de Astrofísica de Andalucía-CSIC, Glorieta de la Astronomía s/n, 18008, Granada, Spain}.
$^{8}${Department of Astronomy, University of Geneva, Chemin d'Ecogia 16, CH-1290 Versoix, Switzerland}.
$^{9}${INFN Sezione di Napoli, Via Cintia, ed. G, 80126 Napoli, Italy}.
$^{10}${INAF - Osservatorio Astronomico di Roma, Via di Frascati 33, 00040, Monteporzio Catone, Italy}.
% ...
}

\acknowledgments 
\tiny{

We acknowledge the financial support of the Spanish MICINN) under contracts PID2019-104114-RB-C32, and PID2021-126536OA-I00 and fellowship PRE2020-093502. A.Dinesh acknowledges the support of a UCM-Santander fellowship.

We would also like to thank the Instituto de Astrof\'{\i}sica de Canarias for the excellent working conditions at the Observatorio del Roque de los Muchachos in La Palma. The financial support of the German BMBF, MPG and HGF; the Italian INFN and INAF; the Swiss National Fund SNF; the grants PID2019-107988GB-C22, PID2022-136828NB-C41, PID2022-137810NB-C22, PID2022-138172NB-C41, PID2022-138172NB-C42, PID2022-138172NB-C43, PID2022-139117NB-C41, PID2022-139117NB-C42, PID2022-139117NB-C43, PID2022-139117NB-C44, CNS2023-144504 funded by the Spanish MCIN/AEI/ 10.13039/501100011033 and "ERDF A way of making Europe; the Indian Department of Atomic Energy; the Japanese ICRR, the University of Tokyo, JSPS, and MEXT; the Bulgarian Ministry of Education and Science, National RI Roadmap Project DO1-400/18.12.2020 and the Academy of Finland grant nr. 320045 is gratefully acknowledged. This work was also been supported by Centros de Excelencia ``Severo Ochoa'' y Unidades ``Mar\'{\i}a de Maeztu'' program of the Spanish MCIN/AEI/ 10.13039/501100011033 (CEX2019-000920-S, CEX2019-000918-M, CEX2021-001131-S) and by the CERCA institution and grants 2021SGR00426 and 2021SGR00773 of the Generalitat de Catalunya; by the Croatian Science Foundation (HrZZ) Project IP-2022-10-4595 and the University of Rijeka Project uniri-prirod-18-48; by the Deutsche Forschungsgemeinschaft (SFB1491) and by the Lamarr-Institute for Machine Learning and Artificial Intelligence; by the Polish Ministry Of Education and Science grant No. 2021/WK/08; and by the Brazilian MCTIC, CNPq and FAPERJ.

We gratefully acknowledge financial support from the following agencies and organisations:
Conselho Nacional de Desenvolvimento Cient\'{\i}fico e Tecnol\'{o}gico (CNPq), Funda\c{c}\~{a}o de Amparo \`{a} Pesquisa do Estado do Rio de Janeiro (FAPERJ), Funda\c{c}\~{a}o de Amparo \`{a} Pesquisa do Estado de S\~{a}o Paulo (FAPESP), Funda\c{c}\~{a}o de Apoio \`{a} Ci\^encia, Tecnologia e Inova\c{c}\~{a}o do Paran\'a - Funda\c{c}\~{a}o Arauc\'aria, Ministry of Science, Technology, Innovations and Communications (MCTIC), Brasil;
Ministry of Education and Science, National RI Roadmap Project DO1-153/28.08.2018, Bulgaria;
Croatian Science Foundation (HrZZ) Project IP-2022-10-4595, Rudjer Boskovic Institute, University of Osijek, University of Rijeka, University of Split, Faculty of Electrical Engineering, Mechanical Engineering and Naval Architecture, University of Zagreb, Faculty of Electrical Engineering and Computing, Croatia;
Ministry of Education, Youth and Sports, MEYS  LM2023047, EU/MEYS CZ.02.1.01/0.0/0.0/16\_013/0001403, CZ.02.1.01/0.0/0.0/18\_046/0016007, CZ.02.1.01/0.0/0.0/16\_019/0000754, CZ.02.01.01/00/22\_008/0004632 and CZ.02.01.01/00/23\_015/0008197 Czech Republic;
CNRS-IN2P3, the French Programme d’investissements d’avenir and the Enigmass Labex, 
This work has been done thanks to the facilities offered by the Univ. Savoie Mont Blanc - CNRS/IN2P3 MUST computing center, France;
Max Planck Society, German Bundesministerium f{\"u}r Bildung und Forschung (Verbundforschung / ErUM), Deutsche Forschungsgemeinschaft (SFBs 876 and 1491), Germany;
Istituto Nazionale di Astrofisica (INAF), Istituto Nazionale di Fisica Nucleare (INFN), Italian Ministry for University and Research (MUR), and the financial support from the European Union -- Next Generation EU under the project IR0000012 - CTA+ (CUP C53C22000430006), announcement N.3264 on 28/12/2021: ``Rafforzamento e creazione di IR nell’ambito del Piano Nazionale di Ripresa e Resilienza (PNRR)'';
ICRR, University of Tokyo, JSPS, MEXT, Japan;
JST SPRING - JPMJSP2108;
Narodowe Centrum Nauki, grant number 2023/50/A/ST9/00254, Poland;
The Spanish groups acknowledge the Spanish Ministry of Science and Innovation and the Spanish Research State Agency (AEI) through the government budget lines
PGE2022/28.06.000X.711.04,
28.06.000X.411.01 and 28.06.000X.711.04 of PGE 2023, 2024 and 2025,
and grants PID2019-104114RB-C31,  PID2019-107847RB-C44, PID2019-105510GB-C31, PID2019-104114RB-C33, PID2019-107847RB-C43, PID2019-107847RB-C42, PID2019-107988GB-C22, PID2021-124581OB-I00, PID2021-125331NB-I00, PID2022-136828NB-C41, PID2022-137810NB-C22, PID2022-138172NB-C41, PID2022-138172NB-C42, PID2022-138172NB-C43, PID2022-139117NB-C41, PID2022-139117NB-C42, PID2022-139117NB-C43, PID2022-139117NB-C44, PID2022-136828NB-C42, PDC2023-145839-I00 funded by the Spanish MCIN/AEI/10.13039/501100011033 and “and by ERDF/EU and NextGenerationEU PRTR; the "Centro de Excelencia Severo Ochoa" program through grants no. CEX2019-000920-S, CEX2020-001007-S, CEX2021-001131-S; the "Unidad de Excelencia Mar\'ia de Maeztu" program through grants no. CEX2019-000918-M, CEX2020-001058-M; the "Ram\'on y Cajal" program through grants RYC2021-032991-I  funded by MICIN/AEI/10.13039/501100011033 and the European Union “NextGenerationEU”/PRTR and RYC2020-028639-I; the "Juan de la Cierva-Incorporaci\'on" program through grant no. IJC2019-040315-I and "Juan de la Cierva-formaci\'on"' through grant JDC2022-049705-I. They also acknowledge the "Atracci\'on de Talento" program of Comunidad de Madrid through grant no. 2019-T2/TIC-12900; the project "Tecnolog\'ias avanzadas para la exploraci\'on del universo y sus componentes" (PR47/21 TAU), funded by Comunidad de Madrid, by the Recovery, Transformation and Resilience Plan from the Spanish State, and by NextGenerationEU from the European Union through the Recovery and Resilience Facility; “MAD4SPACE: Desarrollo de tecnolog\'ias habilitadoras para estudios del espacio en la Comunidad de Madrid" (TEC-2024/TEC-182) project funded by Comunidad de Madrid; the La Caixa Banking Foundation, grant no. LCF/BQ/PI21/11830030; Junta de Andaluc\'ia under Plan Complementario de I+D+I (Ref. AST22\_0001) and Plan Andaluz de Investigaci\'on, Desarrollo e Innovaci\'on as research group FQM-322; Project ref. AST22\_00001\_9 with funding from NextGenerationEU funds; the “Ministerio de Ciencia, Innovaci\'on y Universidades”  and its “Plan de Recuperaci\'on, Transformaci\'on y Resiliencia”; “Consejer\'ia de Universidad, Investigaci\'on e Innovaci\'on” of the regional government of Andaluc\'ia and “Consejo Superior de Investigaciones Cient\'ificas”, Grant CNS2023-144504 funded by MICIU/AEI/10.13039/501100011033 and by the European Union NextGenerationEU/PRTR,  the European Union's Recovery and Resilience Facility-Next Generation, in the framework of the General Invitation of the Spanish Government’s public business entity Red.es to participate in talent attraction and retention programmes within Investment 4 of Component 19 of the Recovery, Transformation and Resilience Plan; Junta de Andaluc\'{\i}a under Plan Complementario de I+D+I (Ref. AST22\_00001), Plan Andaluz de Investigaci\'on, Desarrollo e Innovación (Ref. FQM-322). ``Programa Operativo de Crecimiento Inteligente" FEDER 2014-2020 (Ref.~ESFRI-2017-IAC-12), Ministerio de Ciencia e Innovaci\'on, 15\% co-financed by Consejer\'ia de Econom\'ia, Industria, Comercio y Conocimiento del Gobierno de Canarias; the "CERCA" program and the grants 2021SGR00426 and 2021SGR00679, all funded by the Generalitat de Catalunya; and the European Union's NextGenerationEU (PRTR-C17.I1). This research used the computing and storage resources provided by the Port d’Informaci\'o Cient\'ifica (PIC) data center.
State Secretariat for Education, Research and Innovation (SERI) and Swiss National Science Foundation (SNSF), Switzerland;
The research leading to these results has received funding from the European Union's Seventh Framework Programme (FP7/2007-2013) under grant agreements No~262053 and No~317446;
This project is receiving funding from the European Union's Horizon 2020 research and innovation programs under agreement No~676134;
ESCAPE - The European Science Cluster of Astronomy \& Particle Physics ESFRI Research Infrastructures has received funding from the European Union’s Horizon 2020 research and innovation programme under Grant Agreement no. 824064.}

\end{document}